\documentclass[final,5p,times,preprint,twocolumn]{elsarticle}

\usepackage{amssymb}
\usepackage{amsmath,subfigure,lipsum,flushend,cuted,booktabs}
\usepackage{color}

\journal{Optics Communications}

\begin{document}

\begin{frontmatter}



\title{Outer scale of the wide-range Prandtl/Schmidt  number spectrum on beam wander for oceanic optical turbulence}

\author[label1]{Jian-Dong Cai}
\author[label1]{Jin-Ren Yao}
\author[label1]{Han-Tao Wang}
\author[label1]{Hua-Jun Zhang}
\author[label1]{Ming-Yuan Ren}
\author[label1]{Yu Zhang\corref{cor1} }
\address[label1]{School of Physics, Harbin Institute of Technology, Harbin 150001, China}
\cortext[cor1]{Corresponding author:zhangyuhitphy@163.com}
\begin{abstract}
Light propagation in ocean is influenced by the refractive-index which is related to temperature, salinity, outer-scale, etc. 
Based on Hill's model 1 (H1), two kinds of oceanic refractive-index spectrum (ORIS) have been proposed to describe the second order characteristic of refractive-index. 
Most recently, several ORIS models were proposed based on Hill’s model 4 (H4), which gave a better precision in high wave-numbers (viscous-diffusive range). 
However, the outer scale, as a key parameter related to practical environment, has not been introduced into any H4-based ORIS.
In this paper, we take the outer-scale parameter into an H4-based spectrum which is adapted to the wide-range Prandtl/Schmidt number [Opt. Express. 20, 11111(2019)].
The proposed outer-scaled spectrum could be used in analyzing wave propagation in limited outer-scaled environment with different values of  average temperature and salinity.
We further derived the beam wander formula of collimated laser beam with the outer-scaled spectrum.
Numerical calculations show that 
the beam wander influenced by outer-scale length $L_{0}$ is more obvious than that influenced by average temperature $\langle T\rangle$, when $L_{0}$  varies from $10 \rm{m}$ to $100 \rm{m}$, and $\langle T\rangle$ ranges from $0^{\circ} \rm C$ to $30^{\circ} \rm{C}$.
When salinity fluctuations prevails ($\omega  \rightarrow 0$), the influence of outer scale becomes weaker.
In contribution proportion of  beam wander, 
the temperature-salt coupling term is much larger than that of temperature or salinity term.
\end{abstract}




\end{frontmatter}


\section{Introduction}
Oceanic refractive-index spectrum (ORIS) plays a key role in oceanic optics \cite{Hill78OSA, Hill2, Nikishov:2000,OK:2019}.
Unlike the atmospheric spectrum governed by temperature and humidity  \cite{Obukhov:1962, Hill78RC,Hill:78,Andrews:1992}, 
ORIS is a 2nd order statistical description affected by the fluctuations of temperature, salinity, and their co-operation \cite{Hill78OSA,Nikishov:2000, Leandri:2011}.
ORIS is well-modified in many cases that based on H1 and H4
 \cite{ Nikishov:2000,Yao:2017,Li:2019,Yi:2018},
but oceanic outer scale, as a potentially important parameter of ORIS, has not been discussed widely.

Indeed, 
the outer scale is a narrow restriction of the range with energy injection \cite{Coulman:1988},
which is defined as the highest degree of anisotropy \cite{Toselli:2014}.
And it affects the properties of light beam obviously, 
like low-frequency behavior \cite{Reinhardt:1972}, beam spread \cite{Yura:1973}, adaptive optics \cite{ V.V95,Ziad:2004}, scintillation index \cite{Andrews:1999,Yi:2012}, and angle of arrival fluctuations\cite{Cui:2014}.
Moreover, the outer scale is directly measured by using the temperature structure function \cite{Clifford:1971}, 
and it is also directly obtained for water turbulence in laboratory conditions \cite{ MD:1997}.
The recent works proposed that using back-reflected light can be considered as a new technique for estimation of atmospheric turbulence outer scale \cite{Kulikov:2019}.
Compared with great significance of the outer scale in atmosphere optics, there are a few reports about oceanic optics \cite{Li:2019}. 

The outer scale has been introduced into H1-based oceanic spectra, which shows a lower beam wander \cite{Yue:2019},
and a lower off-axis scintillation index \cite{Li:2019} for Gaussian beams.
However, as stated in Ref.\cite{Hill2}, 
H4-based spectra are more accurate than H1-based spectra in high wavenumbers. 
Motivated by this, an H4-based oceanic refractive-index spectrum was given, which shows a good precision for wide-range Prandtl/Schmidt number \cite{Yao:2019}.
And an application of the spectrum in a bi-LIDAR system elucidated that the underwater optical turbulence degenerates the spectral density and the degree of coherence \cite{OK:2019bi}.
Consequently,
considering the significance of outer scale and the advantage of recently proposed H4-based spectrum, we introduce an outer-scale parameter into the H4-based spectrum,
and use it to calculate the light propagation in ocean.

This paper is arranged as follows. 
In section 2.1 we develop an outer-scaled oceanic spectrum based on the approximate H4-based spectrum;
Section 2.2 gives the analytical expression of beam wander with the outer-scaled oceanic spectrum.
Section 3 shows a further numerical calculation and discussion.
Finally, Section 4 makes a summary.

\section{Theory}
Power spectrum of oceanic refractive-index fluctuations is generally given by a linear combination of temperature spectrum, salinity spectrum and their co-spectrum.
Each of the spectra is based on H1 \cite{Nikishov:2000} or H4\cite{Yao:2019}.
The H4-based ORIS can denote the complex oceanic environment affected by seasonal or extreme average temperature, salinity and/or other factors.
In this section, we will introduce the outer scale into H4-based spectrum, and derive the beam wander based on it.
\subsection{The modification of ORIS about outer scale}

Generally,  outer scale can be described in three forms\cite{V.Vouter:1995}: 
Exponential form, $\kappa^{-11 / 3}\left[1-\exp \left(-\kappa^{2} / \kappa_{0}^{2}\right)\right]$; 
von K\'arm\'an form, ${\left( {{\kappa ^2} + \kappa _0^2} \right)^{ - 11/6}}$; 
and Greenwood form, ${\left( {{\kappa ^2} + \kappa {\kappa _{0}}} \right)^{ - 11/6}}$,
where $\kappa$ is spatial frequency, 
 ${\kappa _0}$ includes the outer scale.
Here we choose the exponential form for its mathematical simplicity and physical reasonability.


The 3D oceanic spectrum ${\Phi _{n}}(\kappa )$  is given by \cite{Yao:2019}:
\begin{align} \label{eq1}
{\Phi _{n}}(\kappa ) = 
{A^2}{\Phi _{\rm{T}}}(\kappa ) + {B^2}{\Phi _{\rm{S}}}(\kappa ) - 
2AB{\Phi _{{\rm{TS}}}}(\kappa ),
\end{align}
where $A$ is the thermal expansion coefficient;
$B$ is saline contraction coefficient,
and we express the three spectra on the right of Eq.(1) with outer-scaled H4-based model:
\begin{align} \label{eq2}
\nonumber
{\Phi _i}(\kappa ) =& \frac{1}{{4\pi }}\beta {\varepsilon ^{ - \frac{1}{3}}}{\kappa ^{{\rm{ - 11/3}}}}\left[{{\rm{1 - exp( - }}\frac{{{\kappa ^2}}}{{\kappa _0^2}}{\rm{)}}} \right]{\chi _i}{g_i}\left( {\kappa \eta } \right),\\
\quad & i \in \{ {\rm{T}},{\rm{S}},{\rm{TS}}\} ,
\end{align}
with
\begin{align} \label{eq3}
\nonumber 
{g_i}\left( {\kappa \eta } \right) =& \left[ {1 + 21.61{{(\kappa \eta )}^{0.61}}c_i^{0.02} - 18.18{{(\kappa \eta )}^{0.55}}c_i^{0.04}} \right]{\rm{ }} \\
&\times \exp \left[ { - 174.90{{(\kappa \eta )}^2}c_i^{0.96}} \right]{\rm{ }},\\
{c_i} = {0.07}&{2^{4/3}}\beta Pr_i^{ - 1},\\
\kappa_0 = \frac{C_0}{L_0},
\end{align}
where $\beta = 0.72 $ is the Obukhov–Corrsin constant; 
$\varepsilon $ is the dissipation rate of turbulent kinetic energy pur unit mass of fluid ,
which varies in the range of $[{10^{ - 10}},{10^{ - 4}}]$ $\rm{m^2}{s^{-3}}$;
$\eta $ is the Kolmogorov microscale length that varies in the range of $[{10^{ - 4}},{10^{ - 2}}]$ $\rm{m}$;
$\chi _i$ is the ensemble-averaged variance dissipation rate;
$P{r_{\rm{S}}} , P{r_{\rm{TS}}} $ are the temperature Prandtl number and salinity Schmidt number, respectively;
the temperature-salinity Prandtl-Schmidt number $P{r_{{\rm{TS}}}} = 2P{r_{\rm{T}}}P{r_{\rm{S}}}{\left( {P{r_{\rm{T}}} + P{r_{\rm{S}}}} \right)^{ - 1}}$;
$L_0$ is the outer scale;
${C_0} \in [2\pi ,8\pi ]$ is the scaling constant, its value depends on the application, and we choose ${C_0} = 4\pi$ here.

\subsection{Theoretical derivation of beam wander}
According to the outer-scaled H4-based ORIS given in the last section, we derive the beam wander as follows.

Beam wander of Gaussian beam is generally expressed as \cite{BOOK1}:
\begin{align} \label{eq4}
\nonumber \left\langle {r_c^2} \right\rangle  = 
 &\frac{4{\pi ^2}{k^2}{W^2}L}{{n}_{0}^{2}}\int_0^1 {\int_0^\infty  \kappa  } {{\Phi_n}(\kappa )}{H_{\rm{LS}}}(\kappa ,\xi ) \\
& \times \left[ {1 - \exp \left( { - \Lambda L{\kappa ^2}{\xi ^2}/k} \right)} \right]{\rm{d}}\xi {\rm{d}}\kappa,
\end{align}
where  
$k = 2\pi {n_0}/\lambda $ is wavenumber, 
which contains refractive index of seawater ${n_0}$ and incident wavelength $\lambda$;
Beam radius,
 $W = {W_0}\sqrt {\Theta _0^2 + \Lambda _0^2} $,
 is associated with the waist of input Gaussian beam ${W_0}$,
${\Theta _0} = 1 - {\bar \Theta _0}$ is the beam curvature parameter at the input plane,
 ${\bar \Theta _0}$ is the complementary parameter; 
$\Lambda  = 2L/(k{W^2})$ and ${\Lambda _0} = 2L/(k{W_0}^2)$ are respectively the Fresnel ratio of beam at the receiver and transmitter;
 $\xi  = 1 - z/L$ is normalized wide-range that carries transmission distance information $L$ and $z$; 
${H_{\rm{LS}}}(\kappa ,\xi )$ is the Gaussian filter function.
To capture the influence of outer scale, 
let filter function hold the components of beam wander, which is generally given by \cite{BOOK1}:
\begin{align} \label{eq5}
{H_{\rm{LS}}}(\kappa ,\xi ) = \exp \left\{ { - {\kappa ^2}W_0^2\left[ {{{\left( {{\Theta _0} + {{\bar \Theta }_0}\xi } \right)}^2} + \Lambda _0^2{{(1 - \xi )}^2}} \right]} \right\}.
\end{align}
The diffraction effects and small-scale effects are neglected \cite{BOOK1},
so we drop the last term and
consider the following approximation
\begin{align} \label{eq6}
1 - \exp \left( { - \Lambda L{\kappa ^2}{\xi ^2}/k} \right) \approx \Lambda L{\kappa ^2}{\xi ^2}/k,\quad L{\kappa ^2}/k \ll 1.
\end{align}
In cooperating with Eqs.(\ref{eq4}) -(\ref{eq6}), we have
\begin{align} \label{eq7}
\left\langle {r_c^2} \right\rangle  = \frac{8{\pi ^2}{L^3}}{{n}_{0}^{2}}\int_0^1 {\int_0^\infty  {{\kappa ^3}} {\xi ^2}} {{\Phi_n}(\kappa )}\exp \left[ { - {\kappa ^2}W_0^2{{({\Theta _0} + {{\bar \Theta }_0}\xi )}^2}} \right]{\rm{d}}\xi {\rm{ d}}\kappa,
\end{align}
where
${\Theta _0}  \ge  0$ implies the collimated, focused and divergent beam cases of optical transmission respectively.
Here, we adopt the outer-scaled H4-based ORIS in Eq.(\ref{eq2}) and substitute it into Eq.(7),
\begin{align} \label{eq8}
\left\langle {r_c^2} \right\rangle  = \frac{2 \pi {L^3}\beta {\varepsilon ^{ - \frac {1} {3}}}{A^2}{\chi _{\rm{T}}}}{{n}_{0}^{2}}\sum\limits_{} {{Y_{i,j}}}.
\end{align}
and
\begin{align} \label{eq9}
\nonumber {Y_{i,j}} = &\int_0^1 {\int_0^\infty  {{\kappa ^{{d_{i,j}} - \frac {2} {3}}}} \left[ {{\rm{1 - exp( - }}\frac{{{\kappa ^2}}}{{\kappa _0^2}}{\rm{)}}} \right]{\xi ^2}} {b_{i,j}} \\
&\exp \left[ { - {\kappa ^2}({a_{i,j}} + W_0^2{{({\Theta _0} + {{\bar \Theta }_0}\xi )}^2})} \right]{\rm{d}}\xi {\rm{ d}}\kappa.
\end{align}
It can be obtained directly from the Eq.(\ref{eq8}) where the beam wander is proportional to ${\varepsilon ^{ - \frac {1} {3}}}$ and $\chi _{\rm{T}}$.
${Y_{i,j}}$ is a third-order matrix with three submatrixes ${a_{i,j}}$,${b_{i,j}}$ and ${d_{i, j}}$.

\begin{flushleft}
	Further, for the case of a collimated beam ( $\Theta_{0}=1$),
\end{flushleft}
\newpage
	\begin{strip}
	\begin{align} \label{eq10}
\nonumber{Y_{i,j}} &= \int_0^1 {\int_0^\infty  {{\kappa ^{d_{i,j} - \frac {2} {3}}}} \left[ {{\rm{1 - exp( - }}\frac{{{\kappa ^2}}}{{\kappa _0^2}}{\rm{)}}} \right]{\xi ^2}} b_{i,j}\exp \left[ { - {\kappa ^2}(a_{i,j} + W_0^2)} \right]{\rm{d}}\xi {\rm{ d}}\kappa\\
&=\frac{1}{6}\left[ {\frac{1}{{{{\left( {{a_{{i,j}}} + W_0 } \right)}^{\frac{1}{6} + \frac{{{d_{{{i,j}}}}}}{2}}}}}{-}\frac{1}{{{{\left( {{a_{{i,j}}} + W_0  + {\kappa_{0} ^{ - 2}}} \right)}^{\frac{1}{6} + \frac{{{d_{{i,j}}}}}{2}}}}}} \right]{b_{{i,j}}}\Gamma \left[ {\frac{1}{6} + \frac{{{d_{{i,j}}}}}{2}} \right].
	\end{align}
\end{strip}
\begin{strip}
	\begin{align} \label{eq11}
\left\{ {{a_{i,j}}} \right\} = 174.9{\eta ^2}\left[ {\begin{array}{*{20}{c}}
	{c_{\rm{T}}^{0.96}}&{c_{\rm{T}}^{0.96}}&{c_{\rm{T}}^{0.96}}\\
	{c_{\rm{S}}^{0.96}}&{c_{\rm{S}}^{0.96}}&{c_{\rm{S}}^{0.96}}\\
	{c_{\rm{TS}}^{0.96}}&{c_{\rm{TS}}^{0.96}}&{c_{\rm{TS}}^{0.96}}
	\end{array}} \right], 
\left\{ {{d_{i,j}}} \right\} = \left[ {\begin{array}{*{20}{c}}
	{\rm{0}}&{0.61}&{0.55}\\
	{\rm{0}}&{0.61}&{0.55}\\
	{\rm{0}}&{0.61}&{0.55}
	\end{array}} \right]
	\end{align}
	\begin{align} \label{eq12}
	\left\{b_{i, j}\right\} = A^{2} \chi_{\rm{T}}  \left[ \begin{array}{ccc}{1} & {21.61 \eta^{0.61} c_{\rm{T}}^{0.02}} & {-18.18 \eta^{0.55} c_{\rm{T}}^{0.04}} \\ 
	{\frac{1}{\omega^{2}} d r} & {21.61 \eta^{0.61} c_{\rm{S}}^{0.02} \frac{1}{\omega^{2}} d r} & {-18.18 \eta^{0.55} c_{\rm{S}}^{0.04} \frac{1}{\omega^{2}} d r} \\ 
	{-\frac{1}{\omega}(1+d r)} & {-21.61 \eta^{0.61} c_{\rm{TS}}^{0.02} \frac{1}{\omega}(1+d r) }&
	{18.18 \eta^{0.55} c_{\rm{TS}}^{0.04} \frac{1}{\omega}(1+d r)}\end{array}\right];
	\end{align}
\end{strip}
\noindent
where,
$\omega  \in \left[ {{\rm{ - 5}}\left. ,{\rm{0}} \right]} \right.$ is the dominant ratio of temperature and salinity. 
On substituting from Eqs.(\ref{eq9})-(\ref{eq12}) into Eq.(\ref{eq8}), beam wander contains a complex dependence of $\omega $, $L_{0}$ and $c_{i}$.
Besides,  $\chi _{\rm{T}}$, $\chi_{\rm{S}}$ and $\chi _{\rm{TS}}$ follow the expressions \cite{Elamassie:17}:
\begin{align} \label{eq13}
{\chi _{\rm{S}}} = \frac{{{A^2}}}{{{\omega ^2}{B^2}}}{\chi _{\rm{T}}}{d_r},\quad {\chi _{\rm{TS}}} = \frac{A}{{2\omega B}}{\chi _{\rm{T}}}\left( {1 + {d_r}} \right).
\end{align}
and
\begin{align} \label{eq14}
{d_r} \approx \left\{ {\begin{array}{*{20}{l}}
	{|\omega {|} + \sqrt{|\omega {|}{{(|\omega | - 1)}}},}&{|\omega | \ge 1}\\
	{1.85|\omega | - 0.85,}&{0.5 \le |\omega | < 1}\\
	{0.15|\omega |,}&{|\omega | < 0.5}
	\end{array}} \right.,
Pr \in [3,3000].
\end{align}

From the mentioned above, 
a outer-scaled ORIS has been given.
We apply it for the transmission of laser beam, 
especially,
an analytical expression of beam wander is derived.
Now then, it will be made a discussion about its factors.

\section{Numerical results and analyses }

In this section, by using the proposed outer-scaled ORIS Eq.(\ref{eq2}) and expression of beam wander in Eq.(\ref{eq9}),
we compare and discuss the beam wander and its proportions influencedc by  temperature, salinity and their coupling term.

In what follows, 
we set 
${n_0} = 1.34$,
$ A = 2.56 \times {10^{ - 4}}{\deg ^{ - 1}}{\rm{l}}$,
$\chi _{\rm{T}} = {10^{ - 5}}{\rm{K}^2}{\rm{s}^{ - 1}}$, 
$\varepsilon  = 1 \times {10^{ - 2}}{\rm{m}^2}{\rm{s}^{ - 3}}$, 
$L = 15\rm{m}$, 
${W_0} = 0.1\rm{m}$.
With the varying of $\langle T\rangle$, we will achieve the different $P{r_{\rm{T}}}$, $P{r_{\rm{S}}}$ and ${\eta }$ in the table 3 \cite{Yao:2019}. 

\begin{figure}
	\centering
	\includegraphics[width = 0.35\textwidth]{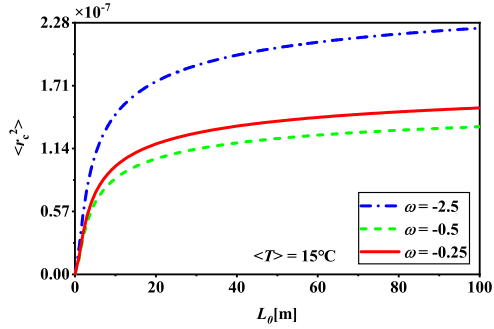}
	\caption{$\langle T\rangle = 15^{\circ} \rm C$, Beam wander versus $ L_0$ with various $\omega$.}
	 \label{fig1}
\end{figure}

\begin{figure}
	\centering
	\includegraphics[width = 0.35\textwidth]{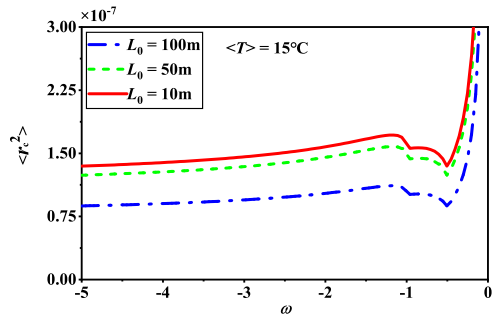}
	\caption{$\langle T\rangle = 15^{\circ} \rm C$, Beam wander versus $\omega$ with various $ L_0$.}
	\label{fig2}
\end{figure}

In Fig.\ref{fig1} we illustrate the influence of outer scale on beam wander when $\langle T\rangle =15^{\circ} \rm C$. 
As the outer scale increases, 
beam wander gradually increases, 
but the extent of  impact gradually slows down. 
Apparently, 
 $\left\langle {r_c^2} \right\rangle_{\omega=-0.25} \geq \left\langle {r_c^2} \right\rangle_{\omega=-2.5} \geq \left\langle {r_c^2} \right\rangle_{\omega=-0.5}$,
there is a nonlinear relationship between $\omega$ and beam wander.
We present details of such a nonlinear relationship in Fig.\ref{fig2}. 
There are three regions: $\omega \in[-5, -1]$ , $[-1, -0.5]$ and $[-0.5, 0]$.
In first region, 
the curve increases slowly, 
then decreases near $\omega = -1$; 
But in second region, it remains stable,
then begins to fall near $\omega = -0.5$; 
Since then, 
the curve rises rapidly, 
and the influence of outer scale converge gradually..
As can be seen from Eq.(\ref{eq14}) , 
the 'jump phenomenon' mainly comes from the modulation of $d_{r}$ given by $\omega $.

\begin{figure} 
	\centering
	\includegraphics[width = 0.35\textwidth]{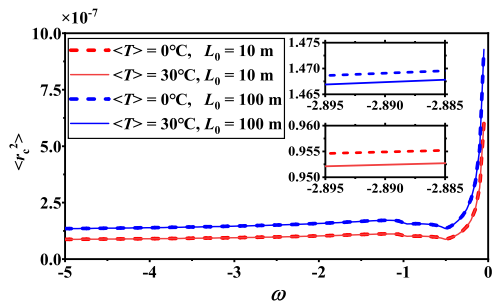}
	\caption{Beam wander versus $\omega$ with various $L_0$ and various $\langle T\rangle$. (a) is local enlarged drawing when ${L_0} = 100\rm{m}$, (b) is local enlarged drawing when ${L_0} = 10\rm{m}$.}
	\label{fig3}
\end{figure}

Figure \ref{fig3} shows the influence of $\langle T\rangle$ and $L_{0}$ on beam wander.
The beam wander influenced by outer scale $L _{0}$ is stronger than that influenced by average temperature, although we set  $L _{0}$ varying from $10 \rm m$ to $100 \rm m$, and $\langle T\rangle$ ranging from $0^{\circ} \rm C$ to $30^{\circ} \rm C$.
From a physical aspect, 
we can understand it as that
the outer scale affects the low frequency inertial region of power spectrum more seriously than average temperature.

\begin{figure*}
	\centering
\subfigure	{
	\includegraphics[width = 0.35\textwidth]{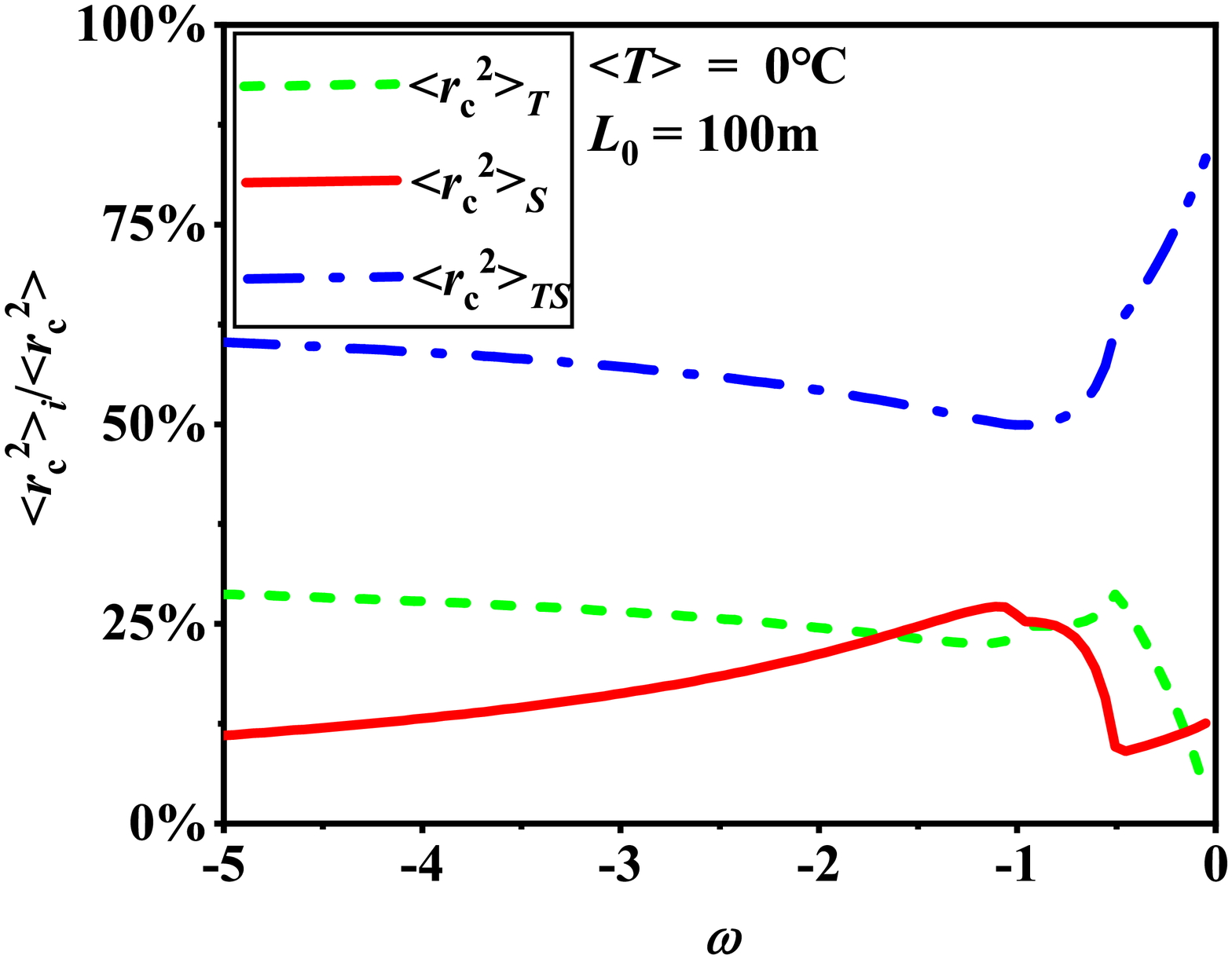}
	\label{a}}
\subfigure	{
	\includegraphics[width = 0.35\textwidth]{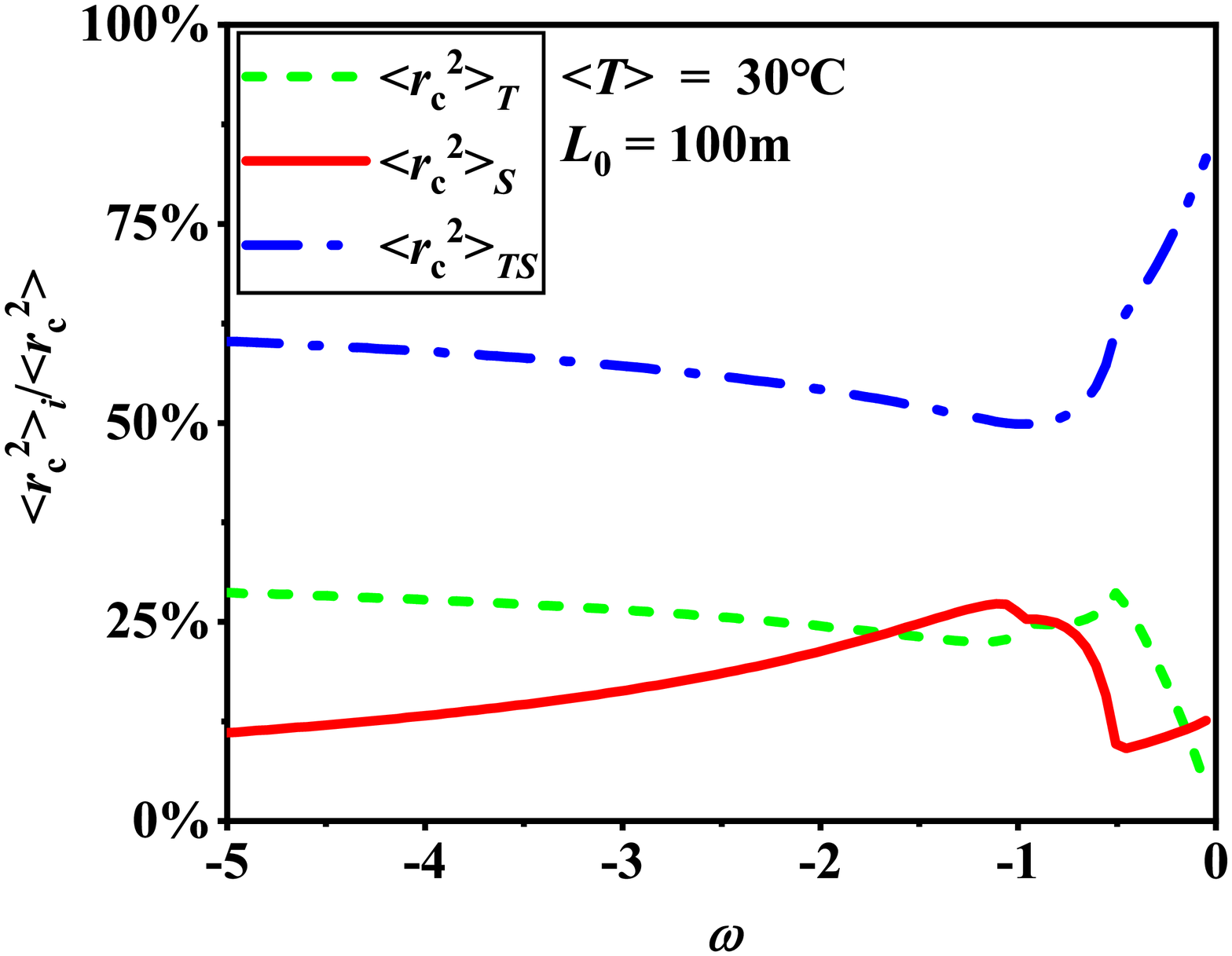}
	\label{b}}
\subfigure	{
		\includegraphics[width = 0.35\textwidth]{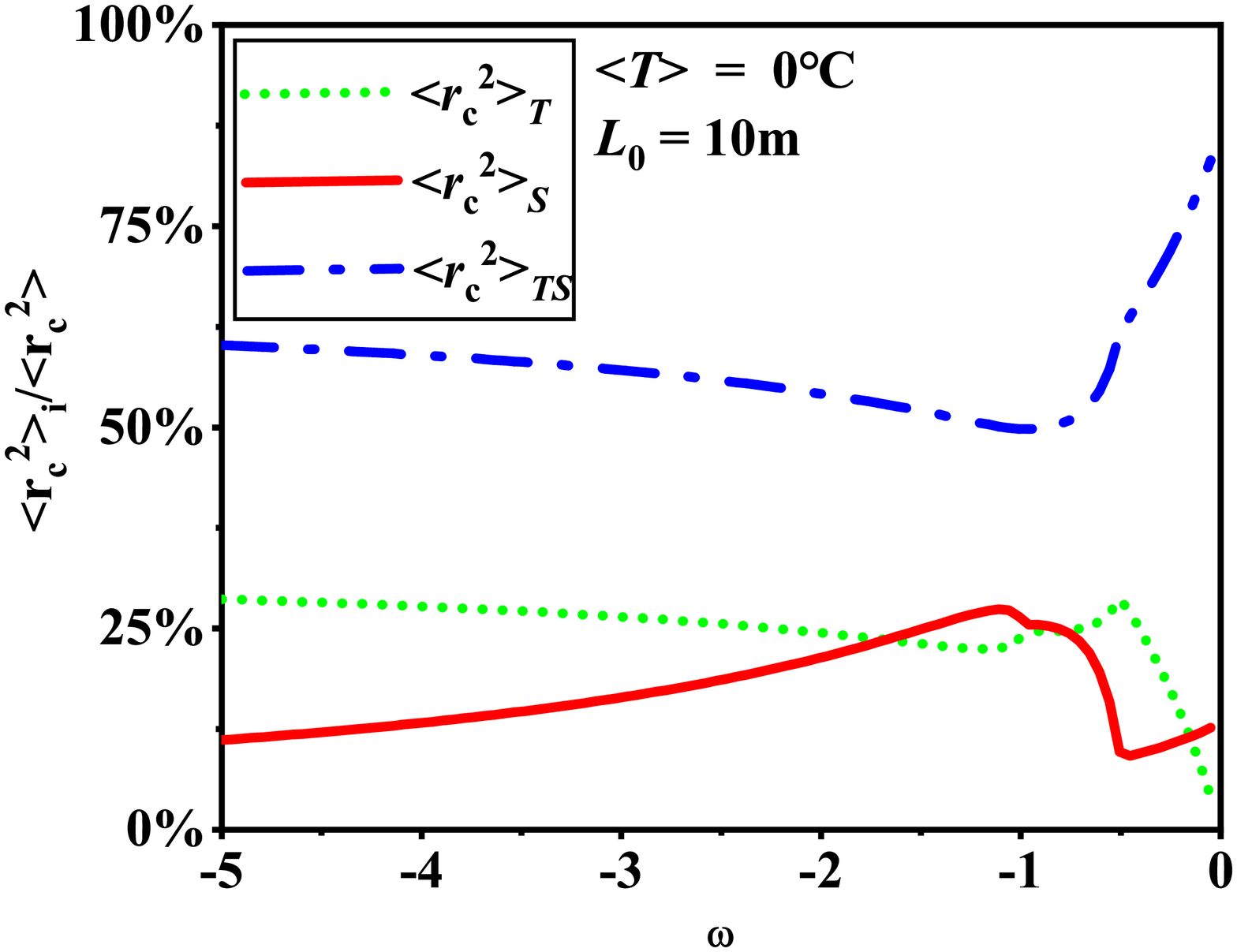}
	\label{c}}
\subfigure	{
		\includegraphics[width =0.35\textwidth]{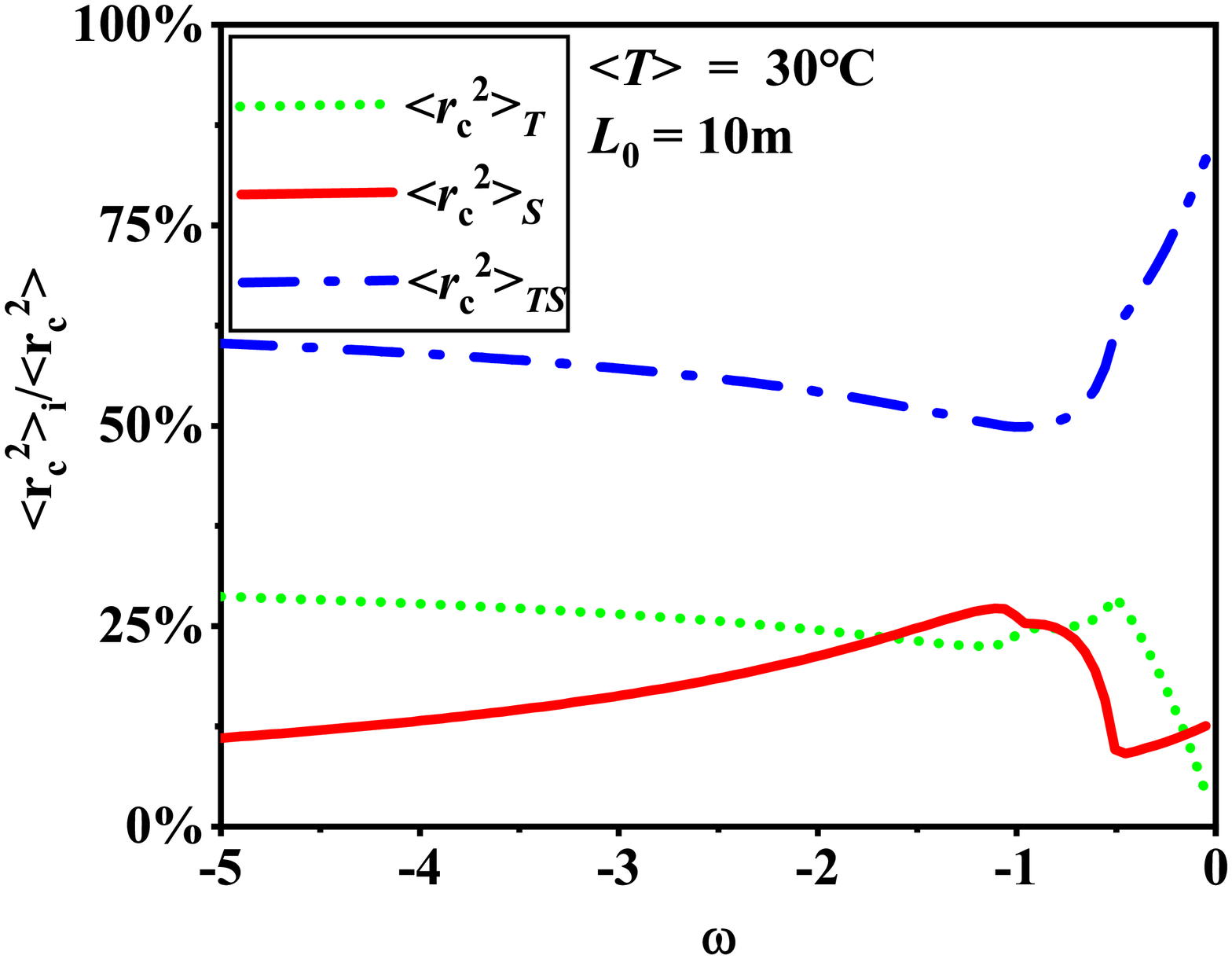}
	\label{d}}
	\caption{Proportion of the beam wander caused by the temperature term, salinity term and the coupling term, for several different values of  $T$ and ${L_0}$:
		(a)  $\langle T\rangle = 0^{\circ} \rm C$, ${L_0} = 10 \rm{m}$, 
		(b) $\langle T\rangle = 30^{\circ} \rm C$, ${L_0} = 10\rm{m}$,
		(c) $\langle T\rangle = 0^{\circ} \rm C$, ${L_0} = 100\rm{m}$, 
		(d) $\langle T\rangle = 30^{\circ} \rm C$, ${L_0} = 100\rm{m}$.}
	\label{fig4} 
\end{figure*}

In addition, it can be seen from Eq.(\ref{eq1}) and Eq.(\ref{eq7})  that the beam wander consists of temperature, salinity, and coupling terms.
The proportions of the three parts varying with $\omega$ are illustrated in Fig.\ref{fig4}(a)(b)(c)(d) for further discussion, and we vary the value of $\langle T\rangle$ and $L_{0}$.
The proportion caused by the coupling term is beyond the others.
It is a slow drop in the first region; For the second region, it remains at the same level;  In the third region, a considerable increase occurs.
This can be interpreted as modulating of the $\chi _i$  in Eq.(\ref{eq13}) due to the relation between $\omega$ and $d_r$,  results in the larger proportion of the coupling term.
And there are almost no changes by relative change between temperature and outer scale, 
i.e. the difference in outer scale and temperature doesn't affect the proportion of beam wander caused by individual terms to the total beam wander obviously.

From the view of  proportion caused by individual terms,
the proportion of  temperature term is decreasing slowly, slowly rising, and sharply falling in the three regions, respectively.
Although  the proportion of  salinity term accounts for a relative small part, its trend is more complex than others. 
For the first region,it slowly rises, then falls a little. In the second region,it rises first, then falls sharply. In the third region, it rises slowly again.
Particularly,
The first region can be considered as the temperature dominated, and the effect of salinity is not very obvious.
In the second region, the proportion of temperature term rises, the proportion of salinity term decreases, 
so the proportion of coupling term is caused  to decrease slowly.
In the third region, the proportion of temperature term drops sharply, and the proportion of salinity term rises slowly, 
but their contributions makes the proportion of coupling term rise sharply.
So we should pay more attention to the change of salinity.

\section{Conclusion}

In conclusion, 
outer scale is introduced into the H4-based oceanic refractive-index spectrum,
which describes the boundary effect in low frequency region. 
Through the outer-scaled H4-based spectrum, we derived the analytical expression of beam wander and plotted the numerical curves of the beam wander varying with several parameters.

Similar with previous report \cite{Yang:2017},
the beam wander rises rapidly with increased outer scale at beginning and then tends to be flat.
When the salinity prevails, the influence of average temperature and outer scale will converge gradually.
We also find that
the beam wander varied with contributed by outer scale is lager than that varied with averaged temperature. 
Moreover, 
for the influence proportion on beam wander,
temperature-salinity coupling term is the largest part.
Although influence of salinity fluctuation on beam wander is relatively small, 
its trend is more complex obviously.
Overall,
The choice of appropriate outer scale influences the information of oceanic optics significantly such as beam wander in communication,  detection and sensing.
It is an important extension with the present theoretical significance of outer scale, 
but of course the experiments in future could extend into parameter regions dependent on outer scale.
we look forward to offering a realistic means to measure results of the fascinating ocean.

\bibliographystyle{unsrt}  
\bibliography{references.bib}

\end{document}